\documentclass[fleqn,usenatbib]{mnras}

\usepackage{newtxtext,newtxmath}

\usepackage[T1]{fontenc}

\DeclareRobustCommand{\VAN}[3]{#2}
\let\VANthebibliography\thebibliography
\def\thebibliography{\DeclareRobustCommand{\VAN}[3]{##3}\VANthebibliography}

\usepackage{graphicx}	
\usepackage{amsmath}	

\newcommand{\Msun}{\mbox{M$_{\odot}$}}
\newcommand{\Rsun}{\mbox{R$_{\odot}$}}

\newcommand{\Mwd}{\mbox{$M_{\mathrm{WD}}$}}

\newcommand{\lppr}{\stackrel{<}{\scriptstyle \sim}}
\newcommand{\lappr}{\raisebox{-0.4ex}{$\lppr$}}
\newcommand{\gppr}{\stackrel{>}{\scriptstyle \sim}}
\newcommand{\gappr}{\raisebox{-0.4ex}{$\gppr$}}

\title[A PCEB with a massive WD \& active companion]{The White Dwarf Binary Pathways Survey - VIII: a post common envelope binary with a massive white dwarf and an active G-type secondary star}

\author[M.-S. Hernandez et al.]{
M.S. Hernandez$^{1,2}$,\thanks{E-mail: mercedes.hernandez@postgrado.uv.cl  matthias.schreiber@usm.cl}
M.R. Schreiber$^{1,2}$,
S.G. Parsons$^{3}$,
B.T. G\"{a}nsicke$^{4,5}$,
O. Toloza$^{1,2}$,
M. Zorotovic$^{6}$\newauthor
R. Raddi$^{7}$,
A. Rebassa-Mansergas$^{7,8}$,
J.J. Ren$^{9}$.
\\
\\
$^{1}$Departamento de F{\'i}sica, Universidad T\'ecnica Federico Santa Mar\'ia, Av. España 1680, Valpara{\'i}so, Chile.\\
$^{2}$Millennium Nucleus for Planet Formation, NPF, Av. España 1680, Valpara{\'i}so, Chile\\
$^{3}$Department of Physics and Astronomy, University of Sheffield, Sheffield S3 7RH, UK.\\
$^{4}$University of Warwick, Department of Physics, Gibbet Hill Road, Coventry, CV4 7AL, UK.\\
$^{5}$Centre for Exoplanets and Habitability, University of Warwick, Coventry CV4 7AL, UK.\\
$^{6}$Instituto de F{\'i}sica y Astronom{\'i}a de la Universidad de Valpara{\'i}so, Av. Gran Breta\~na 1111, Valpara{\'i}so, Chile.\\
$^{7}$Departament de F{\'i}sica, Universitat Polit{\`e}cnica de Catalunya, c/Esteve Terrades 5, E-08860 Castelldefels, Spain.\\
$^{8}$Institute for Space Studies of Catalonia, c/Gran Capit{\`a} 2-4, Edif. Nexus 201, 08034 Barcelona, Spain.\\
$^{9}$Key Laboratory of Space Astronomy and Technology, National Astronomical Observatories, Chinese Academy of Sciences,\\ Beijing 100101, P. R. China.\\
}

\date{ Accepted 2022 September 30. Received 2022 September 29; in original form 2022 June 29 }

\pubyear{2015}

\begin{document}
\label{firstpage}
\pagerange{\pageref{firstpage}--\pageref{lastpage}}
\maketitle

\begin{abstract}
The white dwarf binary pathways survey is dedicated to studying the origin and evolution of binaries containing a white dwarf and an intermediate-mass secondary star of the spectral type A, F, G, or K (WD+AFGK). 
Here we present CPD-65\,264, a new post common envelope binary with an orbital period of 1.37\,days that contains a massive white dwarf {\bf ($ 0.86\pm 0.06\,\mathrm{M}_{\odot}$)} and an intermediate-mass ($1.00\pm0.05\,\mathrm{M}_{\odot}$) main-sequence secondary star.
We characterized the secondary star and measured the orbital period using high-resolution optical spectroscopy. The white dwarf parameters are determined from {\it HST} spectroscopy. In addition, {\it TESS} observations revealed that up to 19 percent of the surface of the secondary is covered with starspots. Small period changes found in the light curve indicate that the secondary is the second example of a G-type secondary star in a post common envelope binary  with latitudinal differential rotation. 
Given the relatively large mass of the white dwarf and the short orbital period,  future mass transfer will be dynamically and thermally stable and the system will evolve into a cataclysmic variable. The formation of the system can be understood assuming common envelope evolution without contributions from energy sources besides orbital energy. CPD-65\,264 is the seventh post common envelope binaries with intermediate-mass secondaries that can be understood assuming a small efficiency in the common envelope energy equation, in agreement with findings for post common envelope binaries with M-dwarf or sub-stellar companions.  
\end{abstract}

\begin{keywords}
binaries: close -- white dwarfs -- solar-type-- stars: activity
\end{keywords}



\section{Introduction}

Close binary stars containing at least one white dwarf are important for a wide variety of astrophysical contexts ranging from understanding the occurrence rates and delay time distributions of SN\,Ia explosions \citep[e.g.][]{mennekensetal10-1} to characterizing the low frequency gravitational  
wave background \citep[e.g.][]{korol17-1}. 
Despite this importance, we still do not fully understand the formation and evolution of these fascinating objects. 

The formation of most 
white dwarf binaries  with periods shorter than a few weeks is thought to be caused by common envelope evolution \citep{Paczynski76,webbink84-1}. Indeed, the distributions of {\bf these} close white dwarf binaries with M-dwarf \citep{zorotovicetal10-1,Camacho14} or substellar companions \citep{Lagosetal21-1,Zorotovic22} can be well reproduced by simple prescriptions of common envelope evolution. 

The situation is more complex when the initial main-sequence binary consists of two stars with masses $\gappr\,1\mathrm{M}_{\odot}$. Such binaries are the progenitors of close white dwarf binaries with secondary stars $\gappr\,1\mathrm{M}_{\odot}$ as well as of close double white dwarfs. It seems that for these populations, models based on common envelope evolution alone are unable to reproduce the characteristics of observed samples \citep[e.g.][]{Nelemans00}. Instead at least two evolutionary channels, i.e. common envelope evolution and stable but non-conservative mass transfer seem to be required to explain the observed populations in both cases \citep{Webbink08, Woods12, Lagos22}. 
In addition, at least one system (IK\,Peg) can only be understood as a post common envelope binary if energy sources in addition to orbital energy contribute during the common envelope phase. Interestingly, IK\,Peg contains a massive white dwarf ($1.19$\Msun) and it has therefore been suggested that if mass transfer starts when a relatively massive donor star is close to the tip of the AGB, recombination energy might play an important role \citep{Rebassa12}. 

However, the currently available observed samples of both close double white dwarfs \citep[e.g.][]{Schreiber22, Napiwotzki20} 
and close white dwarf with F or G type companions \citep{Lagos22} are rather small and/or heavily affected by selection effects.  In particular, IK\,Peg remains the only systems containing a relatively massive white dwarf (exceeding $0.8\,\Msun$) and it therefore remains unclear whether other energy sources during common envelope evolution need to be considered for all post common envelope binaries that contain massive white dwarfs or if IK\,Peg is perhaps just an outlier and the overall population of post common envelope binaries can be understood considering only orbital energy during common envelope evolution.
To progress with this situation, we are currently performing a large scale survey of 
white dwarfs with close intermediate mass companions $\gappr1\,\mathrm{M}_{\odot}$ {\it The White Dwarf Binary Pathways Survey}. 

Finding white dwarfs in close binary systems is relatively simple if the companion is of a sub-stellar class or a low-mass main-sequence (spectral type M) star  \citep{alberto10, alberto16}. However, when the companion to the white dwarf is of spectral type A, F, G or  early-K, the latter completely outshines the white dwarf at optical wavelengths which makes finding these objects in spectroscopic surveys difficult. To overcome 
this problem, we combined optical 
 \cite[e.g. The Radial Velocity Experiment- RAVE,][]{kordopatisetal13-1}
and ultraviolet observations  \citep[Galaxy Evolution Explorer-{\it GALEX},][]{Bianchi14}, to select main-sequence stars with an excess at ultraviolet wavelengths which is indicative for the presence of a white dwarf companion star \citep{parsons16,Rebassa-Mansergas17}.


Here we present a detailed characterization of CPD-65\,264, a white dwarf with a G-type secondary star in a close orbit (1.37\,days). 
The system can be understood as a post common envelope binary that did not require any additional energy (apart from orbital energy) to expel the envelope of the giant and that simply will evolve into a cataclysmic variable in the future. This system increases the number of known post common envelope binaries with intermediate-mass secondaries to seven \citep[][]{parsons15, Hernandez21, Hernandez22}. 
 Among the systems discovered by our survey, CPD-65\,264 contains the most massive white dwarf ($0.86$\,\Msun). 
The large mass of the white dwarf in CPD-65\,264 implies that the progenitor of the white dwarf had evolved to late stages on the AGB when the mass transfer started that led to common envelope evolution. As shown by \citet[][their figure 6]{Rebassa12}, post common envelope binaries containing white dwarf masses exceeding $\sim0.8$\,\Msun are the most suitable targets to test common envelope theories because already at relatively short periods ($\lappr\,3-4\,$days depending somewhat on the secondary star mass) these systems would provide 
evidence for extra energy sources contributing to common envelope evolution. The fact that the period of CPD-65\,264 is well below this threshold further indicates that in the vast majority of cases orbital energy is sufficient to explain the observed properties of post common envelope binaries. Perhaps, only for post common envelope binaries with 
very large white dwarf masses, exceeding $1\,\Msun$ such as the white dwarf in IK\,Peg, recombination energy becomes important. 

Using the available {\it Transiting Exoplanet Survey Satellite} \citep[{\it TESS},][]{Ricker15} data, we also find the secondary star to be differentially rotating and very active (spots cover 8 to 19 per cent of its surface). 
As we have observed similar patterns in two post common envelope binaries previously characterized by our survey, differential rotation seems to rather frequently occur in rapidly rotating G-type stars. {\it TESS} light curves of post common envelope binaries can therefore be used to study activity in the fast rotation regime.

\section{Observations}

We used optical high-resolution spectroscopy to determine the orbital period of the system and to characterize the secondary star. {\it HST} far-ultraviolet spectroscopy was used to measure the white dwarf parameters. In what follows we briefly describe the observational set-ups and data reduction tools that we utilized to study CPD-65\,264.  

\subsection{High-resolution optical spectroscopy} 

We carried out time-resolved high-resolution optical spectroscopic follow-up observations to confirm the close binarity of CPD-65\,264 by measuring the radial velocity variations.
We used the Ultraviolet and Visual Echelle Spectrograph \citep[UVES,][]{Dekker00} on the ESO-VLT and the Fiber-fed Extended Range Optical Spectrograph \citep[FEROS,][]{Kaufer99} at the 2.2\,m-MPG telescope.
The observations carried out with UVES have a spectral resolution of 58\,000 for a 0.7--arcsec slit. With its two-arms, UVES covers the wavelength range of $3000$--$5000$~\AA\, (blue) and $4200$--$11\,000$~\AA\, (red), centered at 3900 and 5640\,\AA\, respectively. Standard data reduction was performed using the specialized pipeline {\sc EsoReflex} workflow \citep{Freudling13}.  
The data obtained with FEROS has a resolution of $R \approx48\,000$ and covers the wavelength range from  $\sim3500$--$9200$\,\AA. The spectra were extracted and analysed with the {\sc ceres} code \citep[][]{jordan14,Brahm17}, an automated pipeline developed to process spectra coming from different instruments in an homogeneous and robust manner following the procedures described in \citet{Marsh89}. 
The instrumental drift in wavelength through the night was corrected with a secondary fiber observing a Th-Ar lamp. 

\subsection{HST spectroscopy} 

With the purpose of confirming the presence of the white dwarf and measuring its mass, we performed  far-ultraviolet spectroscopic observations with the Space Telescope Imaging Spectrograph \citep[STIS,][]{Kimble98} on-board of the {\it HST}. The observation was carried out on 2021 April  21 as part of the program N$^{\mathrm{o}}~$16224 over a single spacecraft orbit resulting in a spectrum with a total exposure time of  2526\,seconds. We used the MAMA detector and the G140L grating providing a spectral resolution between 960--1440 over the wavelength range of 1150--1730\,\AA. The far-ultraviolet spectrum was extracted and wavelength calibrated following the standard procedures on the STIS pipeline \citep{Sohn19}.



\section{Binary and stellar parameters}

In this section we describe how we use the above described observations to determine the binary and stellar parameters of the system.

\subsection{Orbital Period} 

The first step to obtain the orbital period is to calculate the  radial velocities from the high-resolution spectra. For the UVES spectra we used the cross-correlation technique against a binary mask representative of a G-type star. The uncertainties in radial velocity were computed using scaling relations \citep[][]{jordan14} with the signal-to-noise ratio and width of the cross-correlation peak, which was calibrated with Monte Carlo simulations. Radial velocities from FEROS spectra were obtained during data processing with the {\sc ceres} code which also calculates radial velocities using cross-correlation. A total of 15 spectra were analyzed, the entire list of measured radial velocities can be found in Table \ref{tab:RVS}.  The statistical uncertainties of the radial velocities derived from the UVES data are slightly larger than those derived from FEROS because the weather conditions were slightly worse which translated to a slightly lower signal to noise ratio.

\begin{table}
    \setlength{\tabcolsep}{4pt}
	\centering
	\caption{CPD-65\,264 radial velocity measurements. The given uncertainties are purely statistical.}
	\label{tab:RVS}
	\begin{tabular}{llll} 
		\hline
		Instrument   &BJD& RV & error\\
		            &   &[$\mathrm{km~s}^{-1}$] & [ $\mathrm{km~s}^{-1}$]\\
		\hline
FEROS & 2457000.73641 & 6.92 & 0.08 \\ 
FEROS & 2457001.73740 & -78.53 & 0.13 \\
FEROS & 2457002.73840 & 55.30 & 0.10 \\
FEROS & 2457003.54079 & 36.24 & 0.08 \\
FEROS & 2457003.62190 & 71.67 & 0.09 \\ 
FEROS & 2457003.67652 & 91.36 & 0.08 \\
FEROS & 2457003.73820 & 107.96 & 0.09 \\
FEROS & 2457003.79036 & 116.33 & 0.09 \\
FEROS & 2457004.53880 & -81.35 & 0.19 \\
FEROS & 2457004.69224 & -55.47 & 0.09 \\
FEROS & 2457004.75222 & -34.29 & 0.10 \\
FEROS & 2457004.80855 & -10.69 & 0.12 \\
FEROS & 2457383.58068 & 103.14 & 0.08 \\
UVES &  2458231.49972 & 63.35 & 0.29 \\
UVES &  2458352.87197 & -56.65	& 0.71 \\
		\hline
	\end{tabular}
\end{table}

\begin{figure}
	\includegraphics[width=\columnwidth]{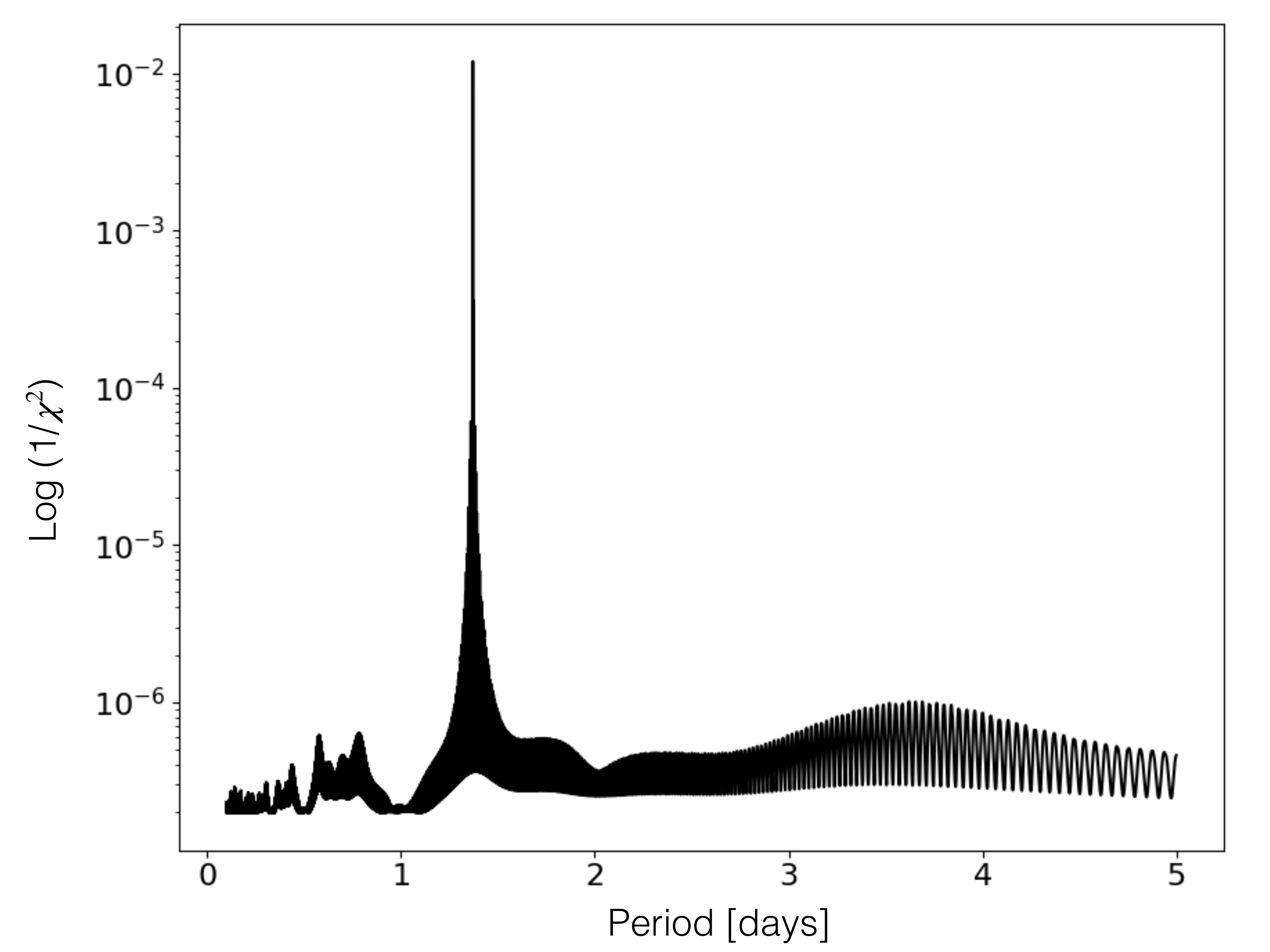}

    \caption{ Periodogram of the radial velocity measurements of CPD-65\,264. The highest peak corresponds to the orbital period of the system. The fit provides a $\chi^2$ of 104.8 and a reduced $\chi^2$ of $8.7$. This relatively large value indicates that systematic errors slightly dominate the purely statistical uncertainties of our radial velocity measurements.}
    \label{fig:RV_periodogram}
\end{figure}

\begin{figure}
	\includegraphics[width=\columnwidth]{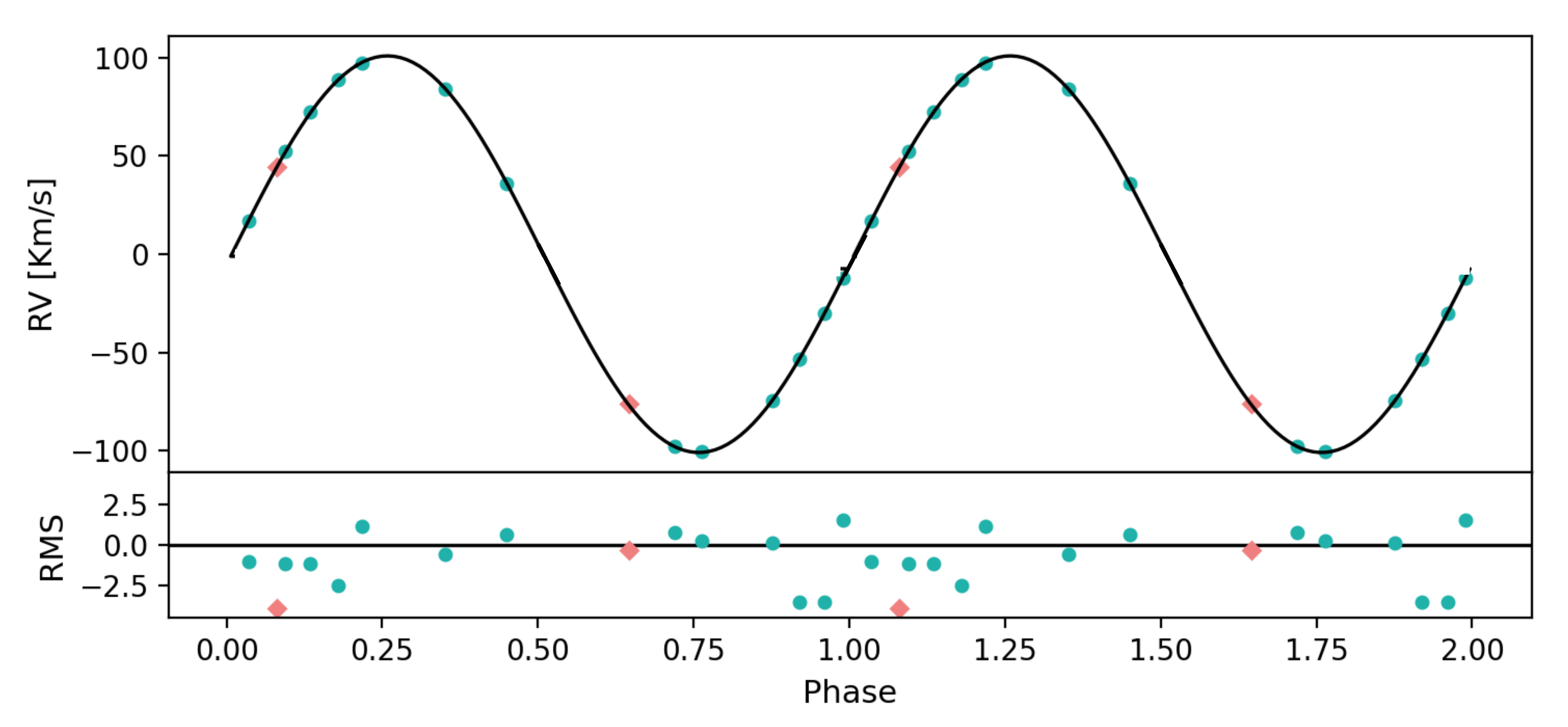}

    \caption{  Radial velocity curve of CPD-65 264 phase folded over the orbital period of 1.37\,days. Radial velocities measurements derived from spectra taken with FEROS are represented with the green dots, while the red diamonds represent radial velocities measured from UVES spectra. The residuals exceed what is expected from the very small statistical uncertainties and indicate that a systematic error of the order of $\sim 1.93$ km/s dominates.  }
    \label{fig:RVC}
\end{figure}

 We used the radial velocities to calculate the orbital period following a least-squares spectral analysis, i.e. we fitted the measurements with a sinusoid of a range of periods and determined $\chi^2$. 
The highest peak (smallest $\chi^2$) in Fig.\ref{fig:RV_periodogram} provides the orbital period of $1.3704$\,days. The phase-folded radial velocity curve is shown Fig.\,\ref{fig:RVC} and clearly illustrates the overall agreement of the fit with the data. 
Inspecting the residuals, however, we note that the scatter of the measurements around the model solution is larger than expected from the statistical uncertainties of our radial velocity measurements. This is in agreement with the relatively large reduced $\chi^2$ of $8.7$ we obtained from the sinosoidal fit and indicates that the statistical errors of our radial velocity measurements significantly underestimate the 
true uncertainties, i.e. systematic errors dominate the radial velocity measurements. We used the scatter around the sinosoidal fit to estimate the systematic error and obtained $1.93$\,km/s (standard deviation). 
As discussed in \citet{parsons15} this systematic radial velocity uncertainty is likely due to
the main-sequence star’s large rotational broadening causing small
systematic errors during the cross-correlation process. For completeness, we note that performing the period determination with the larger systematic uncertainties leads to exactly the same results (except of an unimportant increase of the uncertainty of the measured orbital period).


\subsection{The secondary star} 

We adopted the method described in \citet{Hernandez22} to measure the stellar parameters of the main-sequence star. The procedure is divided in two steps. First we determined the initial values for effective temperature ($T\mathrm{_{eff}}$), surface gravity ($\log{g}$), metallicity ($Z$) and rotational broadening ($v\sin{i}$), by normalizing one of the FEROS spectra and fit it with MARCS.GES\footnote{https://marcs.astro.uu.se/} models \citep{Gustafsson08} using iSpec \citep{Blanco2014}.  
\begin{figure}
    \centering
	\includegraphics[width=\columnwidth]{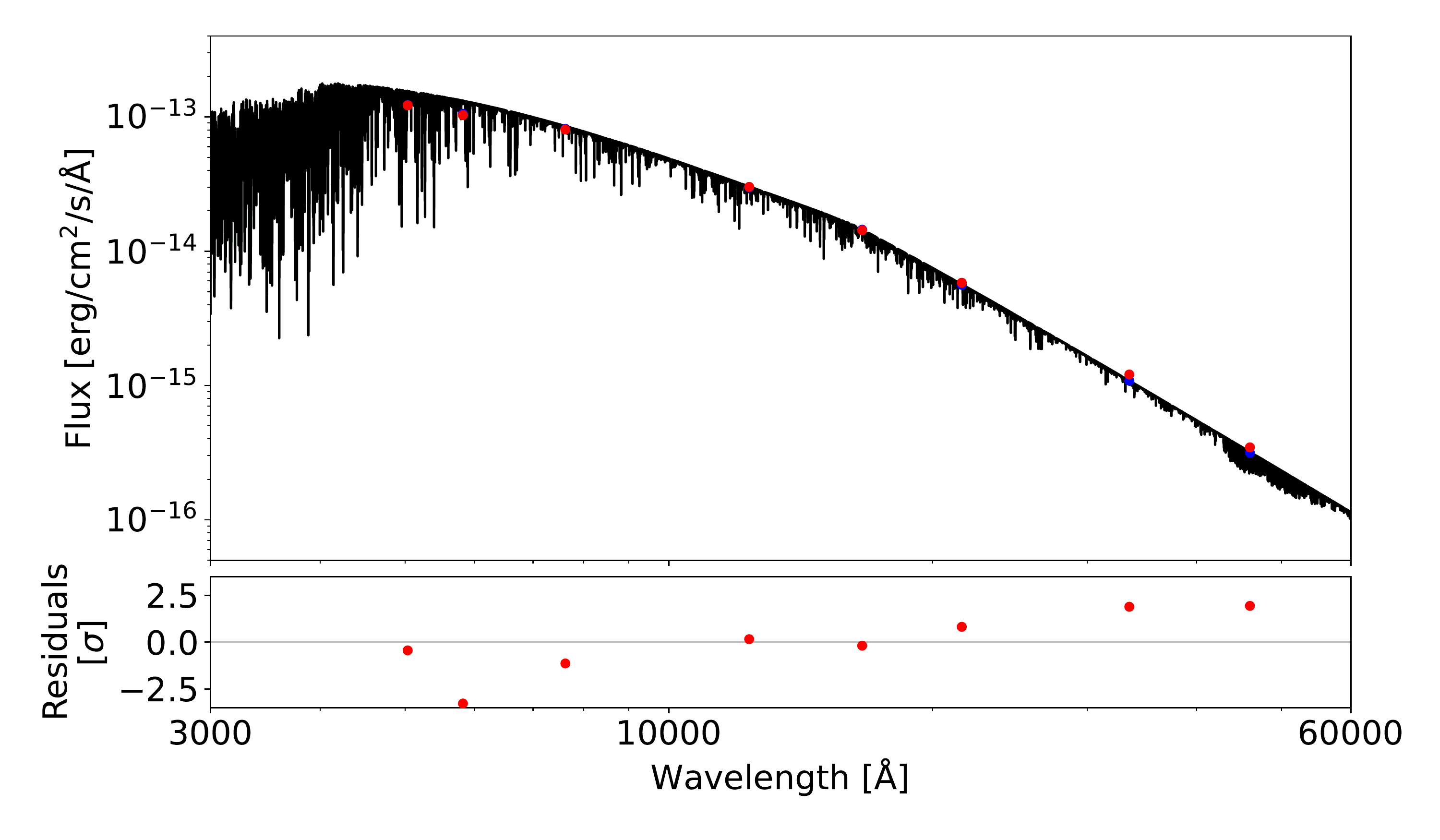} 
	\caption{ Spectral energy distributions created from $G_\mathrm{BP}$, $G_\mathrm{RP}$ and $G$ {\it Gaia} bands, $J$, $H$ and $K_s$ band data from {\rm{2MASS}}, and $W1$ and $W2$ band data from WISE (red dots) were fitted with MARCS.GES theoretical spectra (black line). 
	}
    \label{fig:SED_MS}
\end{figure}
We started off this spectral fit at the values delivered by the {\sc ceres} pipeline, i.e. $T\mathrm{_{eff}}$=5800\,K, $\log{g}$=4.5\,dex, $v\sin{i}$=50\,km\,s$^{-1}$ and solar metallicity, but also performed fits with the initial values perturbed by $T\mathrm{_{eff}}\pm$100\,K, $\log{g}\pm$0.5\,dex, $v\sin{i}\pm$10\,km\,s$^{-1}$, $Z\pm$0.5\,dex. The procedure always converged to the same best fit.


Second, we created a spectral energy distribution (SED) using the {\it Gaia} EDR3 \citep{Gaia20} $G_\mathrm{BP}$, $G_\mathrm{RP}$ and $G$ magnitudes, along with $J$, $H$ and $Ks$ band data from the Two Micron All-Sky Survey  \citep[{\rm{2MASS}},][]{Cutri03}, and $W1$ and $W2$ band data from Wide-field Infrared Survey Explorer   \citep[{\em WISE}][]{Cutri12} and complemented this information with the parallax (4.85$\pm$0.01\,mas) and reddening (E(B-V)=0.060$\pm$0.005\,mag) which are provided by {\it Gaia} EDR3 and the STILISM reddening map  \footnote{https://stilism.obspm.fr/}  \citep{Lallement19, Capitanio17}, respectively.
We then fitted the SED taking into account reddening and parallax using the 
Markov Chain Monte Carlo (MCMC) method \citep{Press07} to determine the final values of mass, radius, effective temperature and surface gravity with their corresponding uncertainties. 
As initial parameters we used the effective temperature and surface gravity previously obtained in step one while the radius was initialized at a value for a main-sequence star with the corresponding $\log{g}$ and $T\mathrm{_{eff}}$. 
The resulting values with their uncertainties are listed in Table \ref{tab:parameters}. The obtained SED 
is presented in Fig.\,\ref{fig:SED_MS}.

\subsection{The white dwarf} 

\begin{figure}
	\includegraphics[width=\columnwidth]{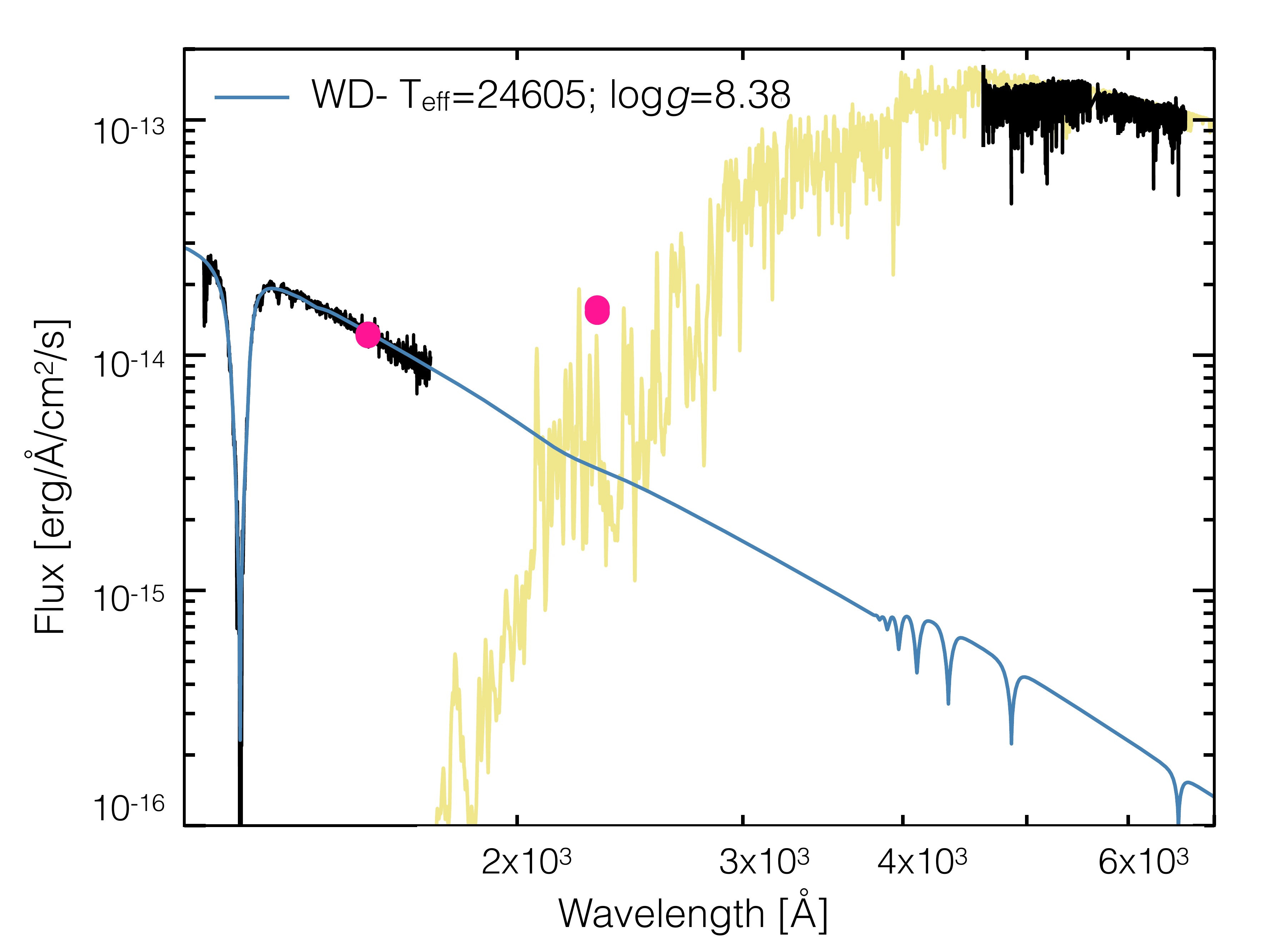}
    \caption{Spectral energy distribution of CPD-65\,264 covering the ultraviolet and optical wavelength range. We show the observed spectra in black including the ultraviolet {\it HST} spectrum of the white dwarf and the optical UVES spectrum of the main-sequence star. In yellow we plot the best MARCS.GES synthetic model of the main-sequence star. The blue line represent the pure-hydrogen synthetic model of the white dwarf. Additionally, we include the {\it GALEX} near- and far-ultraviolet fluxes as pink dots. The combined flux from the WD models and the main sequence star model underestimate the {\it Galex} near-ultraviolet flux. We assume that the model of the G-type secondary star is underpredicting the near-ultraviolet emission as steady chromospheric emission is not taken into account but likely occurring in active G-type stars. We observed the same effect in previously studied systems \citep{Hernandez22}. }
    \label{fig:WD_SED}
\end{figure}

To obtain the white dwarf mass, radius, effective temperature and surface gravity, we fitted the \textit{HST}/STIS spectrum of CPD-65\,264 to a synthetic spectrum of a pure hydrogen atmosphere white dwarf \citep{koester10-1}. 
To that end, we created a grid of synthetic spectra where the effective temperatures spread from 12\,000--30\,000\,K spaced by steps of 200\,K and the surface gravity spans over the range of 6.0-9.0 divided into steps of \,0.1, and establishing the mixing length parameter to 0.8. 
We used the MCMC code provided by the {\sc emcee} python package \citep{Foreman13}, assuming for the reddening and parallax the same priors as for the secondary star (we show the best fit of the spectra in Fig. \ref{fig:WD_SED}).
This procedure provides the surface gravity and the effective temperature of the white dwarf. 

To obtain the white dwarf mass and radius we 
interpolated cooling models from \citep{bedard20}.
The cut-off in the last 100 steps of the chain allowed us to get the best values of the mass and radius which is nearly independent of the assumed thickness of the atmosphere.
We used the marginalized distribution to find the white dwarf parameters and their statistical errors  ($\log{g}=8.381\pm0.005$\,dex, $T_{\mathrm{eff}}= 24\,605\pm48$\,K, \Mwd =$0.865\pm0.003\,\Msun$, $R=0.01004\pm0.00004\,\Rsun$ and cooling age of $7.859\pm0.005$\,Myr).  Furthermore, using the binary mass-function, and the data obtained so far, we deduce the inclination of the system, which is indicated in Table\,\ref{tab:parameters}.  

 We note that the above uncertainties of the white dwarf parameters are purely statistical, i.e. systematic errors are not included. The true uncertainties are certainly larger. \citet[][]{Barstow03} compared $\log\,g$ and $T_{\mathrm{eff}}$ derived from analysing the Balmer lines with those deduced from observations of the Ly$_{\alpha}$ line and obtained a relatively large scatter. We very roughly estimate the true uncertainty of the white dwarf mass by assuming an increased uncertainty of $0.1$\,dex in $\log\,g$ and $800$\,K in $T_{\mathrm{eff}}$ which is broadly consistent with the scatter in \citet[][their figures 9 and 10]{Barstow03}. With this estimate we should be on the safe side 
given that \citet{gianninasetal11-1} estimate smaller systematic uncertainties typically around $1.5$ per cent in $T_{\mathrm{eff}}$ and $0.04$\,dex in $\log(g)$ (albeit from fitting optical data). 
The assumed systematic uncertainties given above translate into more realistic uncertainties of the white dwarf parameters which are listed in Table\,\ref{tab:parameters}.



%


\

\begin{table}
	\centering
	\caption{Summary of the binary and individual stellar parameters of the CPD-65 264.  All uncertainties are purely 
	statistical except of those of the white dwarf parameters which are rough estimates of the dominating systematic errors. 
	}
	\label{tab:parameters}
	\begin{tabular}{ll}
		\hline
           Parameter &  CPD-65 264\\
          		\hline
           m$_{{\it V}}$ [mag] & 11.17 $\pm$ 0.09\\
           Orbital Period [days] & 1.3704 $\pm 0.0001 $   \\
            Phase zero [BJD] &  2457004.874 $\pm$ 0.007    \\
           {\it a} [R$_{\odot}$] & 6.44 $\pm$ 0.01\\
           Distance [pc] &  206.01 $\pm $ 0.47  \\
           Inclination [deg] &  64$\pm$ 1    \\
           Sec. Amplitude [km\,s$^{-1}$] & 100.83 $\pm $ 0.09   \\
           Sec. $v\sin{i}$ [km\,s$^{-1}$] & 38.0 $\pm $ 2.0  \\
           $V_\gamma$ [km\,s$^{-1}$]&  19.19 $\pm$ 0.10   \\
           E[$B-V$][mag] & 0.060 $\pm$ 0.005    \\
           Sec. log g [dex] & 4.39 $\pm0.02 $    \\
           Sec. Z [dex] &  -0.14 $\pm 0.05$   \\
           Sec. $T_{\mathrm{eff}}$ [K] & 5950 $\pm $ 30  \\
           Sec. Radii [R$_{\odot}$] & 1.06 $\pm $0.01 \\
           Sec. Mass [M$_{\odot}$] & 1.0 $\pm $ 0.05  \\
           WD Mass [M$_{\odot}$] & 0.87$\pm$ 0.06  \\
           WD $T_{\mathrm{eff}}$ [K] & 24605 $\pm$  800  \\
           WD log g [dex] & 8.4 $\pm$  0.1  \\
           WD Radii [R$_{\odot}$] & 0.01004 $\pm$ 0.0008  \\
            WD Cooling Age [Myrs] &  7.86$\pm$ 0.14\\
		\hline
	\end{tabular}
	
\end{table}


\section{The active secondary and the TESS light curve}

Main sequence stars in close binaries tend to rotate faster than single stars
\citep{Avallone22, alberto13}.  In post common envelope binaries the orbit is circular \citep[e.g.][]{Nebot11} and tidal forces should quickly synchronize the rotational and orbital period of the secondary star. For periods below 5 days synchronization should take less than $\sim15$\,Myr \citep[][their figure 4]{Fleming19}. As we shall see, the short orbital period of CPD-65 264 resulted in synchronized rotation despite the young age of the white dwarf ($7.9$\,Myr) which further confirms that the synchronisation time scale decreases for shorter  orbital periods \citep[as expected e.g. from figure 4 of][]{Fleming19}. 

To investigate the rotation of the secondary star in CPD-65 264, we inspected the high-cadence {\it TESS} light curves, which we downloaded from the Mikulski Archive for Space Telescopes (MAST\footnote{https://mast.stsci.edu}) web service.
The star was observed in five sectors (hereafter S02, S03, S04, S07, and S11), whose relevant time spans are listed in Table\,\ref{tab:TESS_log}.

%

%
\begin{table*}
    \caption{Period, amplitude, and starspot coverage derived from the data of the five {\it TESS} sectors. }
    \setlength{\tabcolsep}{8pt}
	\centering
    \begin{tabular}{lllllll}
    \hline
         Sector& Time range & Period &  Normalized  & Spot area & Spot surface\\
         &  [BJD-2457000] & [days] &  amplitude & $[\mathrm{m}^2]$ & [$\%$]\\
    \hline

    Original &&&&\\
    S02 & 1354-1381 &1.4188$\pm$ 0.0001 & 0.0139 $\pm$ 0.0005 &  $5.988 \times 10^{16}$ &11\\
    S03 & 1385-1406 &1.4188$\pm$ 0.0004 &0.0131 $\pm$ 0.0001 &  $5.639 \times 10^{16}$ &10\\
    S04 & 1410-1436  &1.3492$\pm$ 0.0002 & 0.0106 $\pm$ 0.0006  &  $4.577 \times 10^{16}$& 09\\
    S07 & 1491-1516 &1.4188$\pm$ 0.0002 & 0.0114 $\pm$ 0.0003 &  $ 4.922 \times 10^{16}$ &19\\
    S11 & 1601-1623 &1.3492$\pm$ 0.0001 & 0.0240$\pm$ 0.0001 &  $1.057 \times 10^{17}$ &08\\
    \hline
        Residuals &&&&\\
       S02 & 1354-1381 &0.6857$\pm$ 0.0007 & 0.0013$\pm$0.0003 \\
    S03 & 1385-1406 &0.6857$\pm$ 0.0004 &  0.0063$\pm$ 0.0004\\ 
    S04 & 1410-1436  &0.6857$\pm$ 0.0003 & 0.0051$\pm$0.0004 \\
    S07 & 1491-1516 &0.6857$\pm$ 0.0005 & 0.0020$\pm$0.0002\\ 
    S11 & 1601-1623 &0.6857$\pm$ 0.0008 & 0.0016$\pm$0.0003 \\ 
    \hline
    \end{tabular}

    \label{tab:TESS_log}
\end{table*}

\begin{figure}
	\includegraphics[width=\columnwidth]{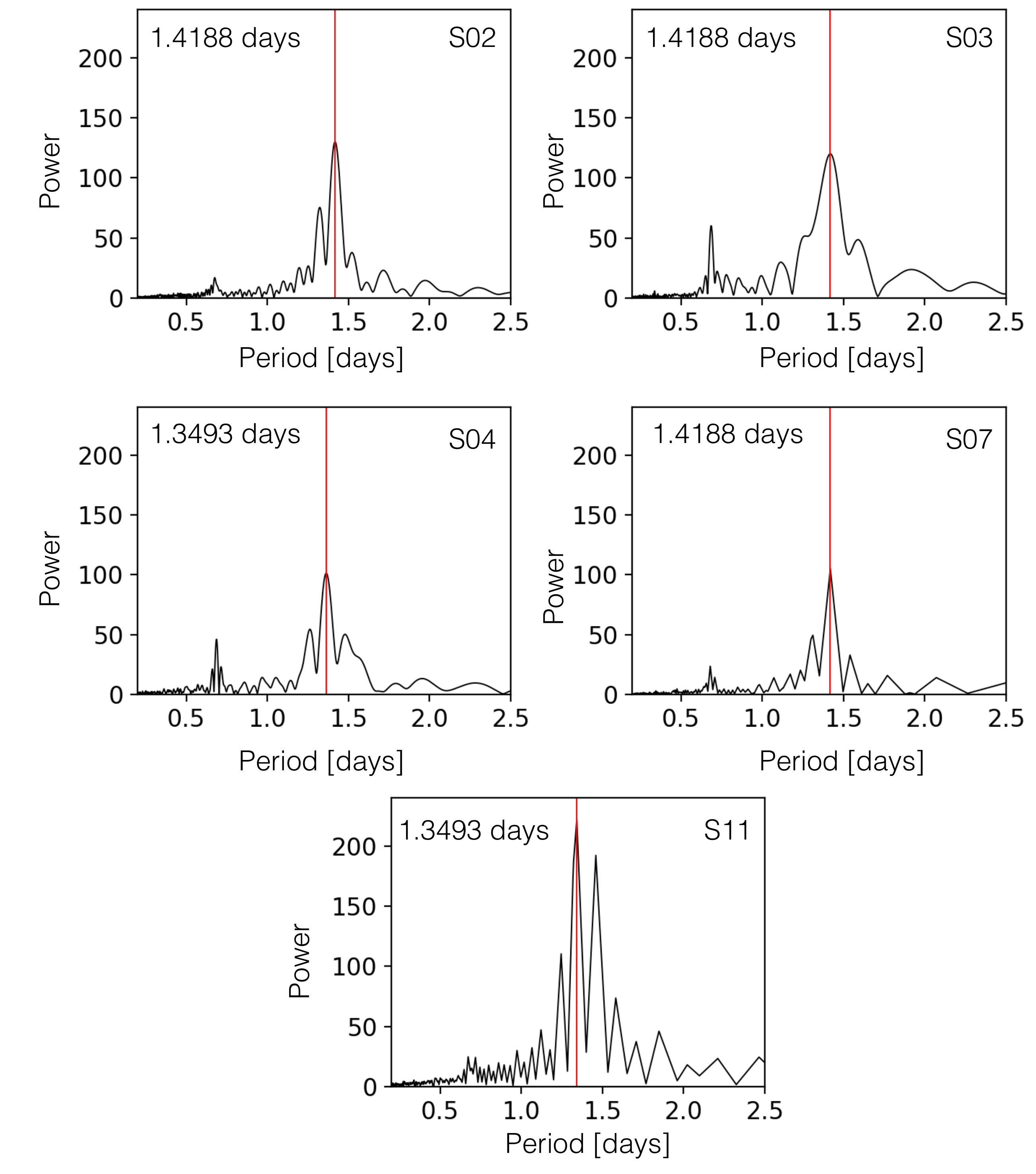}

    \caption{Resulting periodograms of the {\it TESS} light curves for each sector. Sectors are specified on the upper-right corner of each panel, the red line shows the main period identified in each sector (see the periods and errors in Table\,\ref{tab:TESS_log}). }
    \label{fig:TESS_power}
\end{figure}

\begin{figure}
	\includegraphics[width=\columnwidth]{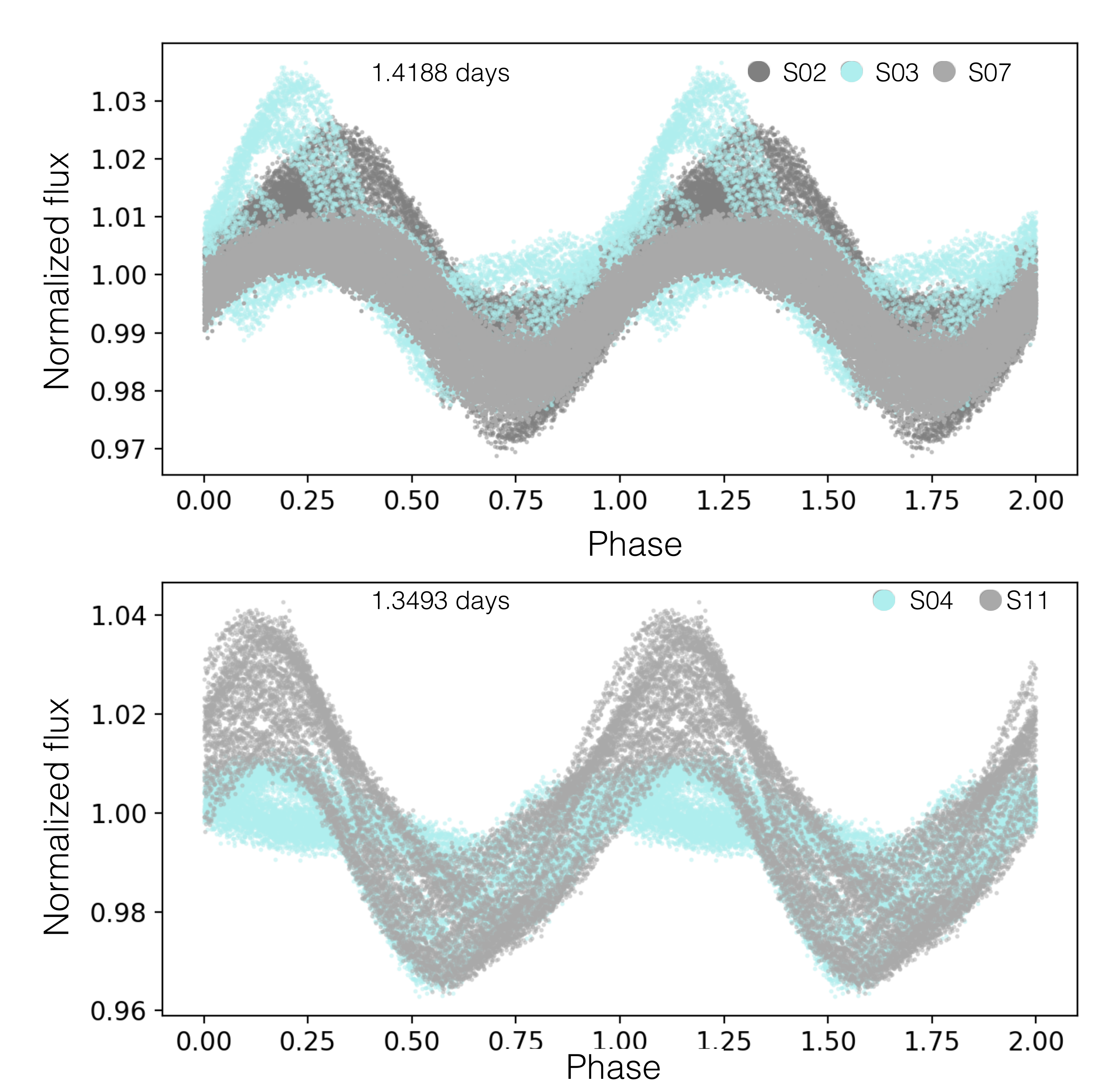}

    \caption{{\it TESS} light curves of CPD-65 264, two different periods were detected in the five sectors. The upper panel shows sectors
    the light curves of sectors S02 (dark gray), S03 (blue), and S07 (light gray) phase folded over 1.4188\,days. The bottom panel show sectors S04 (blue) and S11 (light gray) phase folded over the period of 1.3494\,days.  All light curves used the same phase zero shown in Table\,\ref{tab:parameters} to align with the radial velocity curve. The disagreement between the two photometric periods and the orbital period suggests that the G-type star is rotating differentially, i.e. the period we measure depends on the latitude of the starspots dominating the flux variations. }
    \label{fig:TESS_LC}
\end{figure}

We extracted the Pre-search Data Conditioned Simple Aperture Photometry (PDCSAP) which removes trends caused by the spacecraft, removed all data points with a nonzero quality flag and all NaN values in each sector. We then analyzed each sector with the least-squares spectral method based on the classical Lomb-Scargle periodogram \citep{Lomb76,Scargle82} to obtain the main period of the photometrical {\it TESS } data in each sector (see Fig.\,\ref{fig:TESS_power}). We found two main periods in the five sectors, 1.4188\,days for sectors S02, S03 and S07 while  the light curve of sectors S04 and S11 fit better with a period of 1.3493\,days. The light curve of each sector phase folded over their corresponding period is shown in Fig.\,\ref{fig:TESS_LC}.
While the photometric periods are similar to the orbital period of the system (1.3701\,days), they differ by $\sim84.5$ and $\sim30$\,min, respectively.

We interpret the photometric periods as being caused by starspots and the small but significant differences to the orbital period as being caused by latitudinal differential rotation, i.e.    
the photometric period we measure depends on the latitude of the starspots.

Following the method described in \citet{Notsu19}, we estimated the temperature of the starspots and the surface area covered by them. This goes as follows. First, to obtain the temperature of the starspots ($T_{s}$) we used
equation 4 of \citet{Notsu19}, which is based on the temperature of the  main-sequence star ($T_{\mathrm{MS}}=5950$\,K): 

\vspace{0.2cm}
%
  $  T_{s}= -3.58e^{-5}\,T_{\mathrm{MS}}^2-0.249\,T_{\mathrm{MS}}+808.0+T_{\mathrm{MS}}$.
%
\vspace{0.2cm}

We then used the resulting starspot temperature ($T_{s}\approx4009$\,K) to calculate the area ($A_{s}$) that the starspots cover on the surface of the main-sequence star based on the variation of the light curve using equation 3 from \citep{Notsu19}:  
\vspace{0.2cm}

  $   A_{s}=  2\pi R_{\mathrm{MS}}^2 \frac{\delta F}{F} \left(1-\left(\frac{T_{s}}{T_{\mathrm{MS}}}\right)^4\right)^{-1}$.
%
\vspace{0.2cm}

Here $\frac{\delta F}{F}$ is the normalized amplitude measured from the phase folded light curve of each sector (see "original" section of table\,\ref{tab:TESS_log}, e.g. amplitude/flux zero), and $R_{\mathrm{MS}}$ is the radius of the star. The total surface covered with starspots varies from  $4.577\times 10^{16}$ to $1.057\times 10^{17}\,m^2$, equivalent to 8-19 per cent of the total surface. 
These values correspond to a relatively small spot coverage for solar-type stars with rotational periods between  $0.24-11.16$\,days which spans from $1\times 10^{15}$ to $1\times 10^{18}\,\mathrm{m}^{2}$ \citep{Doyle20}. 

\begin{figure*}
	\includegraphics[width=2.0\columnwidth]{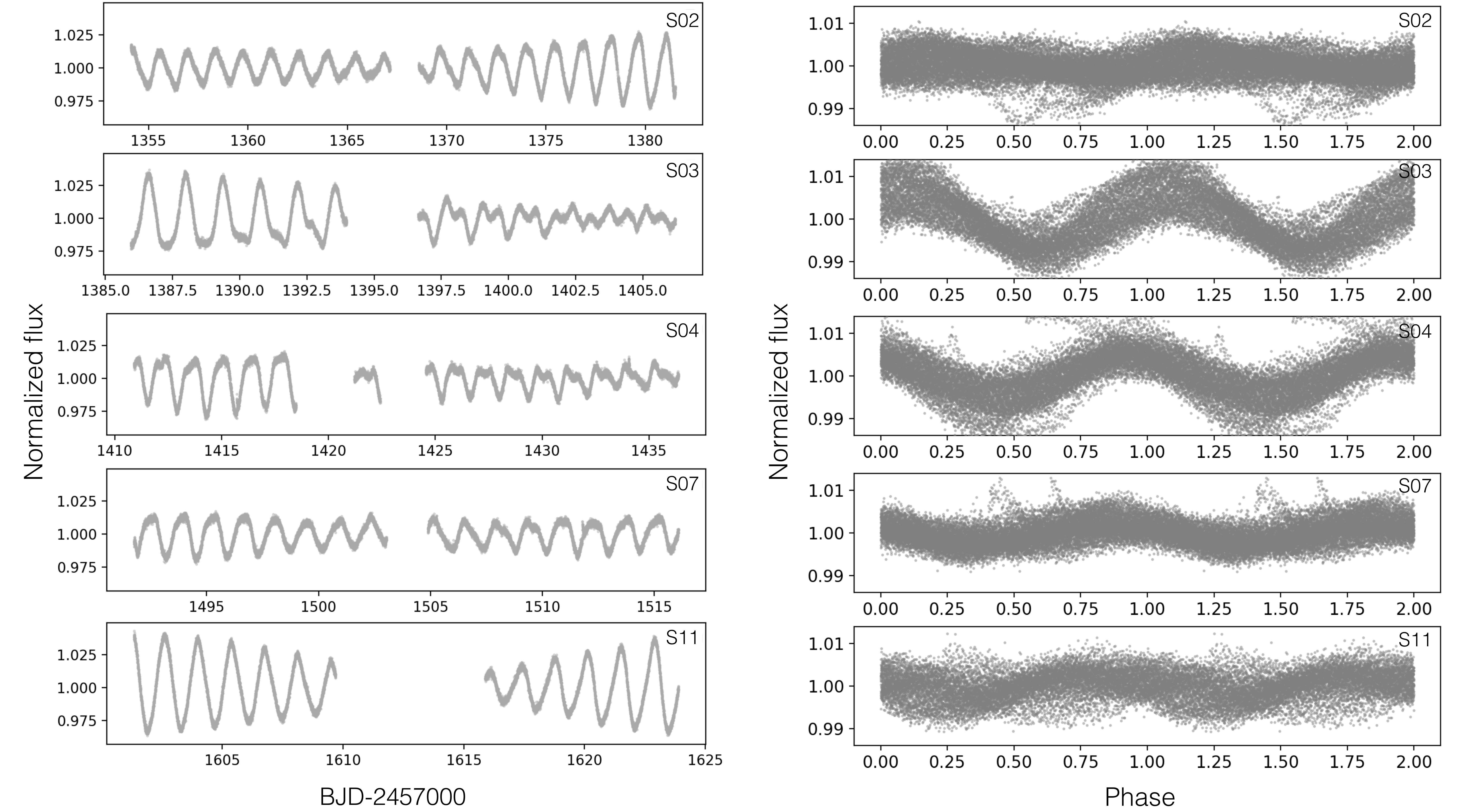}
    \caption{ {\it Left}: The {\it TESS} light curve of CPD-65 264 detailing the magnetic variability of the main-sequence star throughout changing spot structures. {\it Right}: Variation detected in the {\it TESS} light curves after removing their main photometrical signal according to each sector. All five sectors are phase folded over the same period (0.6857\,days) corresponding to the half spectroscopic period  and with the same phase zero (Table\,\ref{tab:parameters}). The variation is consistent with arising from ellipsoidal variations in Sectors S02, S07, and S011 while in the other two sectors contributions from star spots at roughly opposite sides of the secondary need to be assumed.  
    Each panel corresponds to the sector marked in the upper-right corner. }
    \label{fig:tess_original}
\end{figure*}

The left panel of Fig.\,\ref{fig:tess_original} shows the unfolded {\it TESS} light curve which illustrates that short term variations in shape and amplitude are present in each sector. This confirms that the area and/or number of starspots on the surface of the main-sequence star significantly vary with time.   
The average area covered by starspots for each sector is given in Table\,\ref{tab:TESS_log}. 

Looking carefully at the periodograms in Fig.\,\ref{fig:TESS_power}, we identify one signal at half orbital period (0.6857\,days) in all five sectors. To further investigate the origin of this periodic signal, we removed the main dominant photometric periods with their alias from each sector and then phase folded the residuals over half the orbital period (0.6857\,days). 

An obvious interpretation for the signal at half the orbital period are ellipsoidal variations. 
According to \citet{Morris93} and \citet{Zucker07} the expected amplitude of ellipsoidal variations can be estimated
using the following equation: 
\begin{equation}
    \frac{\delta F}{F}=0.15 \frac{(15+u_\mathrm{MS})(1+\beta_\mathrm{MS})}{(3-u_\mathrm{MS})}\left(\frac{R_\mathrm{MS}}{a}\right)^3 q\sin^2{i},
    \label{eq:dff}
\end{equation}
where $\frac{\delta F}{F}$ is the fractional semi-amplitude of the ellipsoidal variation, R$_{\mathrm{MS}}$ is the main sequence star radius, $a$ the semi-major axis,  $q=M_{WD}/M_{MS}$ the mass ratio and $i$ the inclination. For CPD-65\,264 these values are given in Table\,\ref{tab:parameters}. The linear limb darkening coefficient ($u_\mathrm{MS}$)  and the gravity darkening exponent ($\beta_\mathrm{MS}$) were obtained from tables 24 and 29\footnote{https://cdsarc.cds.unistra.fr/viz-bin/cat/J/A+A/600/A30\#/browse} reported by \citet{Claret17}.

The amplitudes predicted by
Eq.\,\ref{eq:dff} range from  $0.00240$ to $0.00372$ 
while the measured normalized amplitude form the residual light curves fluctuate between  $0.00132-0.00641$ in the five sectors (specific values for each sector can be found in "residuals" section of Table\,\ref{tab:TESS_log}).  While sectors S02, S07 and S11 are in a good agreement with the theoretical prediction, in sectors S03 and S04 the amplitude exceeds what is expected from ellipsoidal variations. 

This difference is likely produced by 
a combination of starspot signals and ellipsoidal variations in sectors S03 and S04. 
The left panel of Fig.\,\ref{fig:tess_original}
shows the (not phase folded) {\it TESS} light curve for the five sectors. 
The second part of the {\it TESS} light curve of sectors S03 and 
S04 are quite irregular and show a double peak which we interpret as being caused by a starspots arising at nearly opposite sides of the star which boosts the signal at half the orbital period (right panel of Fig.\,\ref{fig:tess_original}) by up to 56 per cent. 


\section{Past and future of CPD-65 264}

With the stellar masses and the orbital period at hand, it is possible to reconstruct the past evolution of the system, thereby providing constraints on theories of close compact binary formation.  

Using our own tool which combines the stellar evolution code \citep[SSE,][]{hurley00} 
with the common envelope energy equations as described in \citet{zorotovicetal10-1} and Roche geometry, we can obtain the range of possible values for the common envelope efficiency ($\alpha_{\mathrm{CE}}$). 
The algorithm is described in detail in \citet[][see their section 4.1]{Hernandez21}. 
We allowed $\alpha_{\mathrm{CE}}$ to take any value in the range of 0 to 1 and assumed that the change in orbital energy is the only source of energy available to expel the envelope. 
\begin{figure}
    \centering
	\includegraphics[width=0.7\columnwidth]{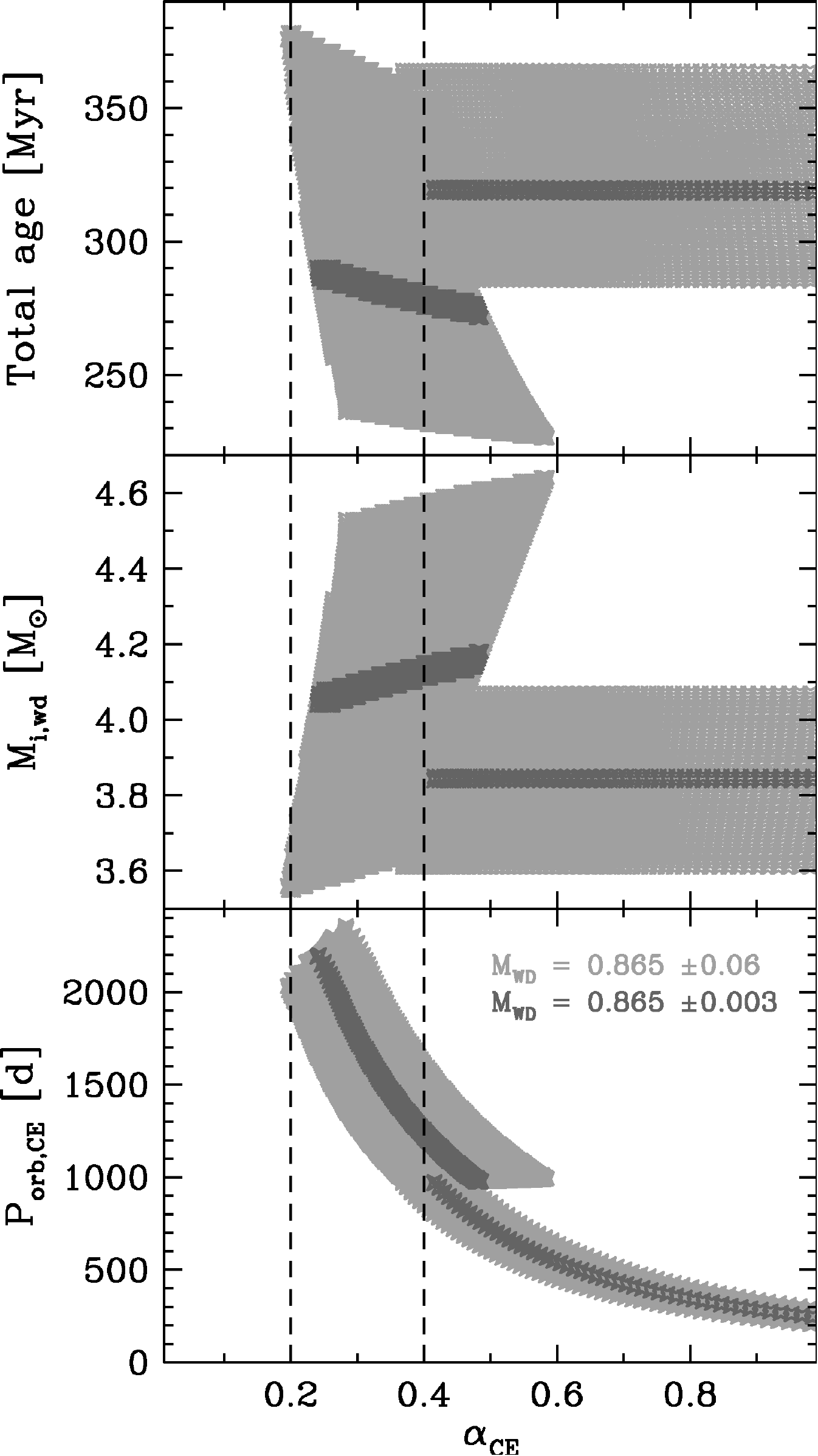}
    \caption{  Reconstruction of the past evolution for CPD-65\,264 and the possible initial configurations of CPD-65\,264 as a function of the common-envelope efficiency $\alpha_{\mathrm{CE}}$.
    Total age of the system (\textit{top}), initial mass of the progenitor of the white dwarf (\textit{middle}), and orbital period at the onset of the common-envelope phase (\textit{bottom}). 
    The dashed lines demarcate the range of  $\alpha_{\mathrm{CE}} = 0.2-0.4$ that works for the reconstruction of all the observed post common envelope binaries with M-dwarf and brown dwarf companions, as well as for the systems previously discovered by our survey.}
    \label{fig:past}
\end{figure}

Figure\,\ref{fig:past} shows, from top to bottom, the allowed solutions for the total age of the system (i.e. time since the binary was born until the common envelope phase + cooling age of the white dwarf), the initial mass of the white dwarf's progenitor, and the period at the onset of common envelope evolution, as a function of the common envelope efficiency. We distinguish those solutions that are consistent with the small error estimated for the white dwarf mass ($0.003\,\Msun$, dark gray) and solutions allowing the error to be slightly larger ($0.06\,\Msun$, light gray). In both cases, we found reasonable solutions with a large range of $\alpha_{\mathrm{CE}}$ and without the need of any extra source of energy, which is consistent with the results we found for all similar systems previously characterized by our survey \citep{parsons15,Hernandez21,Hernandez22}. The breaks observed in the solutions correspond to possible progenitors on different evolutionary stages. Smaller values of $\alpha_{\mathrm{CE}}\simeq0.2-0.4$, consistent with the results obtained for post common envelope binaries with M-dwarf and brown dwarf companions
\citep{zorotovicetal10-1,Zorotovic22}, imply a more massive progenitor ($\sim4.1\,\Msun$) that evolved faster and filled its Roche lobe on the thermally pulsating asymptotic giant branch, where the envelope is more extended and therefore less bound. The initial orbital period in this case should have been larger than $\sim1000$\,days. On the other hand, for larger efficiencies ($\alpha_{\mathrm{CE}}\simeq0.4-1.0$) the progenitor should have been slightly less massive ($\sim3.8\,\Msun$) and filled its Roche lobe on the early asymptotic giant branch, in a binary with initial orbital period in the range of $\sim210-1000$\,days.

\begin{figure}
    \centering
	\includegraphics[width=1.0\columnwidth]{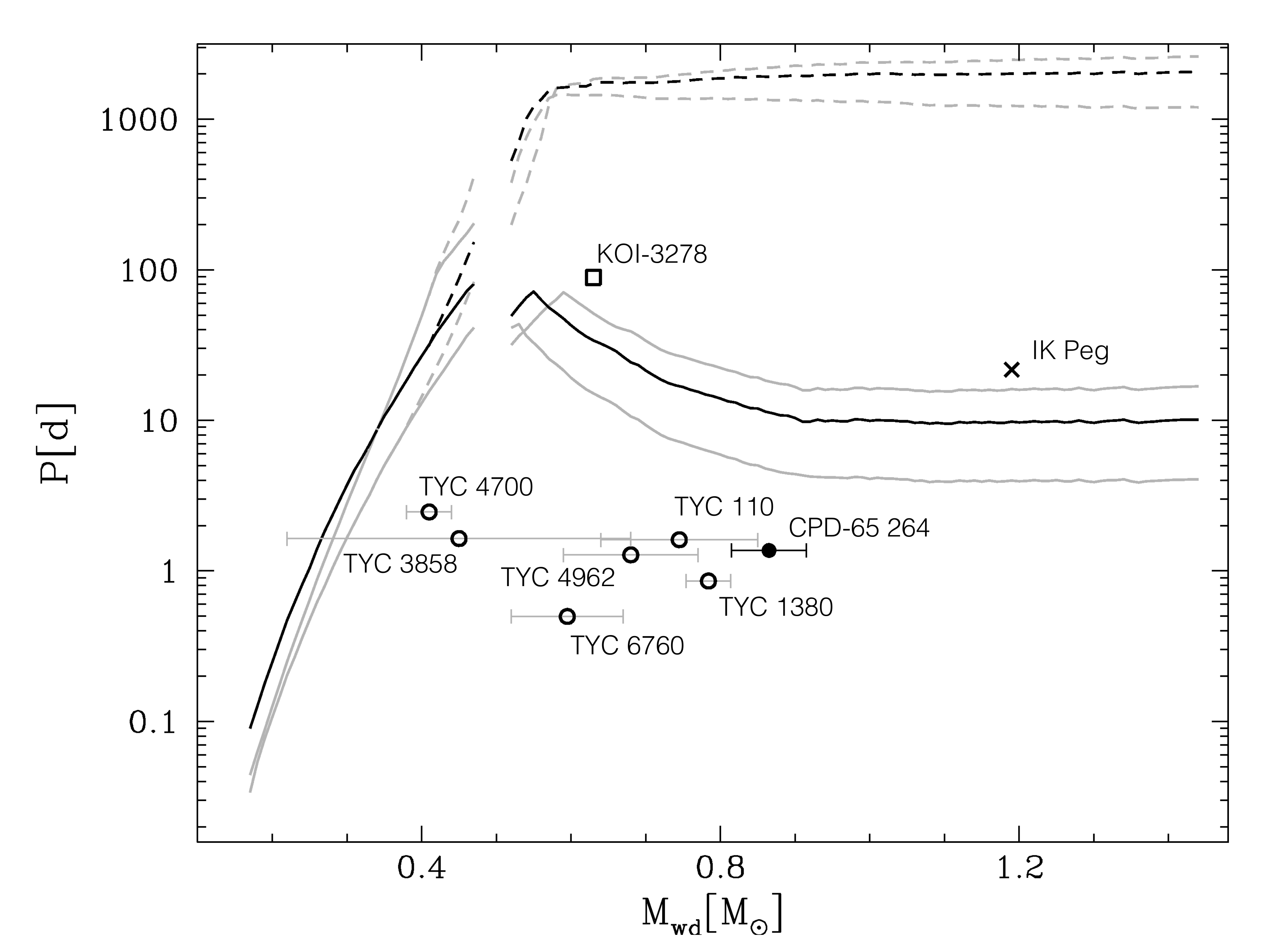}
    \caption{Orbital period limits at which common envelope evolution requires including additional energy sources (solid lines). The limits have been calculated as in \citet{Rebassa-Mansergas12} but assuming larger secondary star masses, i.e. 0.5 (lower grey), 1.0 (black), and 1.5 (upper grey) \Msun. The dashed lines indicate the maximum periods if all the available recombination energy contributes to expelling the envelope. 
    IK\,Peg (the cross) remains the only system whose formation can only be explained by assuming additional energy (e.g. recombination energy) to contribute during common envelope evolution. A second one might be KOI-3278 (square) but it is less clear that this system formed through common envelope evolution. 
    CPD-65\,264 is the system with the second largest white dwarf mass but is clearly separated from the critical period that requires additional energy to contribute. This further indicates that the outcome of common envelope evolution can in the vast majority of cases be well understood by considering only orbital energy. }
    \label{fig:past2}
\end{figure}

 In any case, the formation of CPD-65\,264 can be fully understood by considering only orbital energy during common envelope evolution and by assuming low common envelope efficiencies of $0.2-0.4$ in agreement with previous findings for post common envelope binaries with lower mass secondary stars \citep{zorotovicetal10-1, Zorotovic22}. This is particularly interesting given the relatively large white dwarf mass of CPD-65\,264 (0.86\,\Msun). To illustrate this, we calculated the maximum orbital period predicted for post common envelope binaries assuming 
that the envelope is expelled only through orbital energy. The procedure is based on the reconstruction algorithm from \citet{zorotovicetal11-1} and described in detail in \citet[][]{Rebassa-Mansergas12}. 
In short, we assumed that the white dwarf mass is equal to the core mass of the
giant progenitor at the onset of mass transfer and that the secondary
star mass remains constant during common envelope evolution. 
%
A grid of stellar evolution tracks calculated with the SSE code from \citet{hurley00} then provides all possible progenitor masses and their radii. 
The latter must have been equal to the Roche radius at the onset
of common envelope evolution which leaves as the remaining free parameters the final orbital period 
and the common envelope efficiency. Assuming the maximum common envelope efficiency then provides the longest possible final orbital
period as a function of white dwarf and secondary mass.

 As shown in Fig.\,\ref{fig:past2}, the identification of post common envelope binaries with orbital periods exceeding $4$\,days and white dwarf masses exceeding $\sim0.8$\,\Msun would provide evidence for additional energy sources to play a role during common envelope evolution. 
The fact that the periods we found so far are well below this period limit, in particular in the case of CPD-65 264, indicates that common envelope evolution
can usually be understood without assuming additional energy sources. IK\,Peg remains the only system where assuming only orbital energy fails to reproduce the system parameters we observe today. 
A second system could be KOI-3278 \citep{zorotovicetal14-1} but its 
lower white dwarf mass and much longer period also move it closer to the 
period-white dwarf mass relation for stable mass transfer \citep{rappaportetal95-1} and it is therefore less clear that this system is indeed a post common envelope binary. 
Assuming KOI-3278 is not a PCEB, Fig.\,\ref{fig:past2} could mean that only in the formation of systems with extremely large white dwarf masses, perhaps exceeding 1\,\Msun, recombination energy becomes important. Alternatively, IK\,Peg might just be an outlier that formed through a different yet to be discovered evolutionary channel.

It is also possible to foresee the future evolution of CPD-65\,264 by performing simulations with the Modules for Experiments in Stellar Astrophysics  \citep[\textsc{mesa},][]{paxton11}. To that end, we assumed as initial parameters those reported in Table\,\ref{tab:parameters}
and the {\it star plus point mass with explicit mass transfer rate} module.  

We assumed non-conservative mass transfer (setting the parameter $\beta =1.0$) to start the simulation as nova eruptions should appear as long as the mass transfer stays below the critical values for stable hydrogen burning.  

\begin{figure}
	\includegraphics[width=\columnwidth]{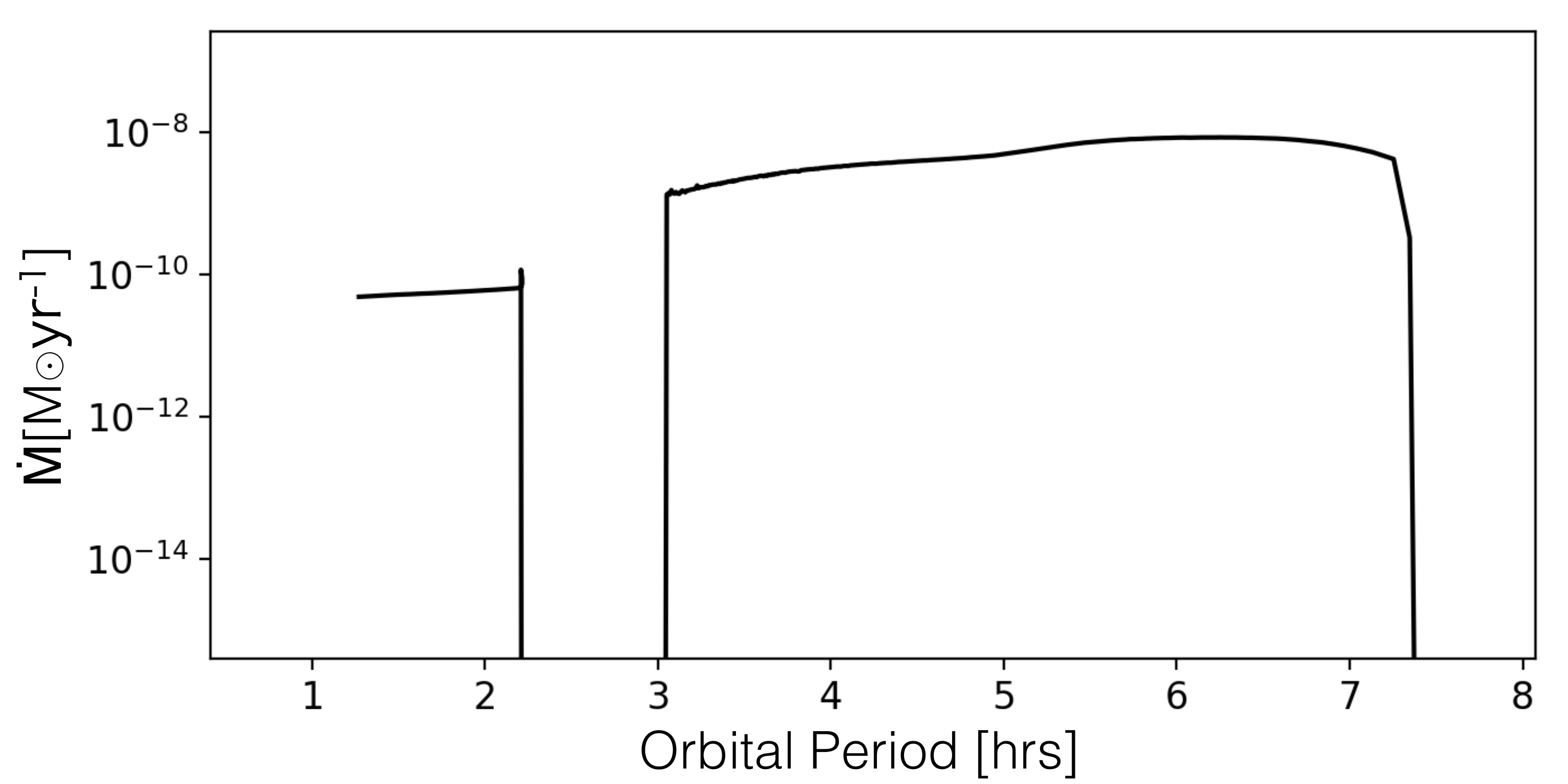}
    \caption{MESA simulation of the future evolution of CPD-65\,264. The initial conditions are a white dwarf mass of 0.86\,M$_{\odot}$ and a main sequence star of 1.0\,M$_{\odot}$ with an orbital period of 1.37\,days. In 1.53\,Gyr years and at an orbital period of 7.7\,hrs, the binary will become a cataclysmic variable, i.e. it will  experience (stable) angular momentum loss driven mass transfer. As the secondary will not substantially evolve, the system will detach when magnetic braking is disrupted and evolve as a detached binary through the orbital period gap. }
    \label{fig:mesa}
\end{figure}

We found that mass transfer should start in 1.53\,Gyr from now at an orbital period of 7.7\,hrs. Throughout its evolution the mass transfer will remain dynamically and thermally stable, the mass of the white dwarf remains constant and the system evolves as a cataclysmic variable.   
The secondary becomes fully convective at a period of 3.0\,hrs, when angular momentum loss will switch from being driven by magnetic breaking to being solely caused by gravitational radiation. 
The binary will reach the minimum orbital period of 1.2\,hrs in 4.78\,Gyrs, when the secondary star has lost 90 per cent of its mass. 
Given the mass of the white dwarf and its small cooling age, the system is currently rather young (see Fig.\,\ref{fig:past}) and therefore the secondary will not evolve off the main sequence and be indistinguishable from cataclysmic variables that descend from binaries with less massive secondary stars. 
The predicted mass transfer rate of CPD-65 264 is shown in Fig.\,\ref{fig:mesa}. 

\section{Conclusions}

We present a detailed characterization of CPD-65\,264, the seventh system that clearly is a post common envelope binary with intermediate-mass secondary \citep[][]{parsons15, Hernandez21, Hernandez22} identified by our survey. We performed optical and {\it HST} spectroscopy to measure the  orbital period and the stellar masses. {\it TESS} photometry 
confirmed the orbital period 
and showed that the secondary star is active and differentially rotating.
We found variations in the {\it TESS} light curve revealing changes in the size and/or number of starspots, covering between 8 to 19 per cent of the effective area of the star that we observe. {\it TESS} light curves of post common envelope binaries can therefore in principle be used to study activity in the fast rotation regime.  


Reconstructing the past and predicting the future evolution of CPD-65\,264, 
we found that the formation of the system can be understood in the context of common envelope evolution without requiring additional energy sources and that in the future the system will become an ordinary cataclysmic variable. 

CPD-65\,264 further indicates that most observed post common envelope binaries can be understood as the outcome of common envelope evolution with no extra energy sources and only a small value of the common envelope efficiency. 
This finding seems to be independent of the mass of the secondary star 
\citep{Lagosetal21-1,Zorotovic22,zorotovicetal10-1}. 

However, apart from short period post common envelope binaries, systems with periods exceeding several months that most likely formed through stable and non-conservative mass transfer have been identified \citep{Kawahara18}, 
and all systems we identified with periods in between a few days and a few months turned out to be contaminants \citep{Lagos22}.
While we cannot yet draw any final conclusions because the sample sizes are too small, it seems that at least two evolutionary channels are required to understand the population of  
close white dwarf binaries with intermediate mass secondary stars.  The recently published data release 3 of the {\it Gaia} mission 
contains a large number of binary stars with measured orbital periods and it is possible to select candidate binaries with compact objects \citep{shahafetal19-1}. Therefore, {\it Gaia} DR3 
may significantly help to identify sufficient numbers of post mass transfer white dwarf binaries with measured periods to provide solid constraints.   

 Carefully analysing this potential of {\it Gaia} DR3 data and eventually
comparing a larger sample with the predictions of binary population models will be important for our understanding of white dwarf binaries in general and potentially help to finally understand evolutionary pathways to SN\,Ia explosions. 
A first step in this direction has been taken by \citep{Korol22} who analysed the unresolved double white dwarf population in {\it Gaia} DR3, a work that should be complemented by studies of their progenitor systems.

\section*{Acknowledgements}

MSH and MRS acknowledge support by ANID, – Millennium Science Initiative Program – NCN19\_171.
MRS and MZ were also supported by FONDECYT (grant 1221059).
SGP acknowledges the support of the STFC Ernest Rutherford Fellowship. 
BTG was supported by the UK STFC grant ST/T000406/1.
OT was supported by a Leverhulme Trust Research Project Grant and FONDECYT grant 3210382.  ARM acknowledges support from Grant RYC-2016-20254 funded by MCIN/AEI/10.13039/501100011033 and by ESF Investing in your future, and from MINECO under the PID2020-117252GB-I00 grant. RR has received funding from the postdoctoral fellowship programme Beatriu de Pin\'os, funded by the Secretary of Universities and Research (Government of Catalonia) and by the Horizon 2020 programme of research and innovation of the European Union under the Maria Sk\l{}odowska-Curie grant agreement No 801370. For the purpose of open access, the author has applied a creative commons attribution (CC BY) licence to any author accepted manuscript version arising.

\section*{Data Availability}

Raw and reduced FEROS and UVES data are available through the ESO archive \url{http://archive.eso.org/cms.html}.



\bibliographystyle{mnras}
\bibliography{TWBPS_VIII} 







\bsp	
\label{lastpage}
\end{document}